\documentclass[12pt]{article}

\usepackage{amsmath}
\usepackage{amssymb}
\usepackage{amsthm}
\usepackage{geometry}
\usepackage{setspace}
\usepackage{hyperref}
\usepackage{graphicx}
\usepackage{booktabs}
\usepackage{natbib}

\newtheorem{proposition}{Proposition}

\title{Beyond the Mean: A Weighted $k$-Sample Omnibus Variance-Ratio
Statistic for Covariate Balance Diagnostics}

{\small
\author{Ariel Linden, DrPH\\
University of California, San Francisco\\
Department of Medicine\\
Division of Clinical Informatics \& Digital Transformation (DoC-IT)\\
San Francisco, CA, USA\\
ariel.linden@ucsf.edu}
}

\date{}
\begin{document}

\maketitle

\section*{Abstract}

Rubin's variance ratio ($VR$) complements the standardized mean
difference by detecting covariate imbalance in spread, but no
purpose-built omnibus extension exists for more than two groups. We
introduce $F_{VR}$, a size-weighted quadratic combination of pairwise
log-variance-ratios that generalizes $VR$ to $k$ groups, together with
geometric-mean and maximum pairwise variance-ratio comparators. $F_{VR}$
has an exact relationship to Rubin's $VR$ at $k=2$ and a mathematical
structure directly parallel to Cohen's $f$. In Monte Carlo simulations
spanning four variance-driven bias mechanisms, $k=3,4,6$, and sample
sizes from 200 to 10{,}000, $F_{VR}$ and the geometric-mean statistic
generally tracked downstream estimation bias at least as well as the
maximum statistic. $F_{VR}$ retained substantially more signal when many
groups shared a very small sample, while absolute-bias thresholds were
unstable at $N=200$. Across mechanisms, $F_{VR}<0.10$ was reasonably
reassuring, values above approximately 0.30 generally indicated concern,
and intermediate values were context-dependent. The statistic is
implemented for arbitrary analysis weights in the Stata command
\texttt{varatio} and illustrated using a multivalued-treatment
application.

\section*{Keywords}

variance ratio; covariate balance; effect size; multivalued treatments;
simulation study

\section{Introduction}

Covariate balance is a necessary step in supporting a causal interpretation
of an estimated treatment effect, achieved by design, as in randomized
experiments, or by adjustment, as in propensity-score weighting
\citep{robins2000}, marginal mean weights through stratification \citep{linden2014}, entropy balancing
\citep{Hainmueller2012}, or matching \citep{lindensamuels2013}. Demonstrating
balance on observed characteristics shows that an adjustment procedure has
balanced the variables it was given; it does not, by itself, guarantee
balance on unmeasured confounders, on which any causal interpretation also
depends \citep{rubin2008}.

For a continuous covariate compared between two groups, balance is
conventionally assessed on (at least) two moments simultaneously: the
standardized mean difference (SMD), and the variance ratio ($VR$)
introduced by Rubin \citep{rubin2001}, who showed that groups can share an
acceptable SMD while differing enough in variance that linear regression
adjustment grossly over- or under-corrects for the remaining bias. That
finding traces to an explicitly empirical source: Rubin's own $VR>2$ (or
$<0.5$) convention is drawn directly from simulation evidence in Cochran
and Rubin \citep{cochranrubin1973}, whose Table~1 --- reproduced in Rubin's
2001 paper --- shows regression-adjustment bias reduction collapsing or
reversing sign once the variance ratio between two groups reaches 2 or its
reciprocal. The threshold marks an empirically located point of
adjustment failure, not an arbitrary round number.

When more than two groups are compared, two very different extensions of
this two-group picture now exist. For the mean, Cohen's $f$ generalizes
the SMD to $k>2$ groups with an exact, derivable relationship to the
two-group case \citep{linden2026cohensf}. For the entire distribution, the
weighted $k$-sample Kolmogorov--Smirnov, Cram\'er--von Mises, and
Anderson--Darling tests extend nonparametric distributional comparison to
$k>2$ groups, again reducing exactly to their two-group counterparts at
$k=2$ \citep{linden2026ksample}. Variance imbalance, however, sits in a
gap between these two extremes: a specific higher moment, more than the
mean captures but far short of the full distribution these nonparametric
tests characterize, and no purpose-built $k$-sample omnibus statistic for
it exists. What is available instead is ad hoc: software such as the R
package \texttt{cobalt} extends Rubin's pairwise $VR$ to multi-category
treatments by reporting the single largest pairwise ratio, following the
``Max2SB'' maximum-aggregation logic developed by Lopez and Gutman
\citep{lopezgutman2017} for standardized mean bias specifically, not for
variance ratios; the extension to variance is an unpublished software
design choice, not an independently validated statistic in its own right.

This paper takes up that gap. We ask the same comparative question already
posed and answered for the mean \citep{linden2026cohensf}: which
$k$-sample statistic best summarizes variance-ratio imbalance --- the
existing maximum-based aggregation, its natural mean-based counterpart,
or a Cohen's-$f$-style quadratic combination? We then ask whether the
answer depends on how imbalance actually translates into downstream
estimation bias, rather than on an abstract target alone. We introduce
$F_{VR}$, a size-weighted quadratic combination of pairwise
log-variance-ratios, alongside two comparators generalized to the same
setting: GMVR, the geometric mean of the pairwise ratios, and the maximum
pairwise ratio itself, the statistic \texttt{cobalt} already reports. All
three are implemented for arbitrary weighted models via the
community-contributed Stata command \texttt{varatio} \citep{linden2026varatio}.
Although much of the motivation above centers on propensity-score
adjustment, the underlying construction (Section~2.1) accepts any
observation-level analysis weights and is not specific to
generalized-propensity-score or inverse-probability of treatment weighting (IPTW): it applies equally to balance assessment after entropy balancing,
calibration weighting, matching weights, or marginal mean weighting
through stratification (MMWS), or any other procedure that produces analysis
weights.

The paper's contributions fall into three categories. The mathematical
contribution is twofold. First, we derive $F_{VR}$'s exact relationship
to Rubin's $VR$ at $k=2$ (Section~2.3). Second, we show that the
minimum-ratio result proven for Cohen's $f$ \citep{linden2026cohensf}
transfers, without modification, to the log-variance scale
(Section~2.5): the same underlying algebra applies to any set of $k$
real numbers, whether they represent group means or log-variances, so
the proof did not need to be redone, only relabeled.
The computational contribution is \texttt{varatio} itself, implementing
$F_{VR}$, GMVR, and the maximum pairwise ratio for arbitrary weighted
models, with a weighted-variance point estimate that remains correct
regardless of whether weights correlate with the covariate being
assessed. The empirical contribution is a Monte Carlo study spanning
four structurally distinct bias-generating mechanisms, sample sizes
from 200 to 10{,}000, and $k\in\{3,4,6\}$, together with a
classification-accuracy analysis that identifies a preliminary,
evidence-based interpretive range for $F_{VR}$, following the same
simulation-derived approach Cochran and Rubin
\citep{cochranrubin1973} originally used to establish the two-group
convention this paper extends.

\section{Methods}

\subsection{Notation and General Framework}

Consider a continuous covariate $X$ and a treatment variable with $k\ge2$
levels, denoted Groups $1,\ldots,k$, with $n_j$ observations in Group $j$
and $N=\sum_j n_j$. Unlike the mean, which is naturally obtained as a
regression coefficient or adjusted predictive margin, variance has no
equally direct model-based analog available across arbitrary model types;
accordingly, each group's variance is computed directly as the weighted
second central moment, normalized by total analysis weight,
\begin{equation}
s_j^2 \;=\; \frac{\sum_i w_{ji}(x_{ji}-\bar x_{j}^{\,w})^2}{\sum_i w_{ji}},
\qquad
\bar x_j^{\,w} \;=\; \frac{\sum_i w_{ji}x_{ji}}{\sum_i w_{ji}},
\end{equation}
using whatever observation-level weights $w_{ji}$ the analysis specifies
--- for instance, inverse-probability or MMWS weights --- and reducing to
the ordinary empirical second central moment (variance with divisor $n$,
not $n-1$) when all weights equal 1. A direct consequence worth noting:
if $X^{*}=aX+b$ for any $a\neq0$, then $s_j^{*2}=a^2s_j^2$, so
$\ell_j^{*}=\log(a^2)+\ell_j$ --- the same additive constant for every
group, which cancels exactly in every deviation $\ell_j-\bar\ell$ used
below. $F_{VR}$, and every pairwise ratio built from it, is therefore
exactly invariant to any location shift or nonzero linear rescaling of
$X$; converting a covariate from dollars to thousands of dollars, or
kilograms to pounds, cannot change the balance diagnosis. We avoid the term
``standard weighted sample variance'' since several competing definitions
exist, particularly regarding finite-sample corrections and effective
degrees of freedom; the quantity above is used as a balance diagnostic,
not as a claim to any particular unbiased-estimator property under a
specific sampling model. This differs from the model-margin approach
used for the mean-based statistics of \citet{linden2026cohensf}. The
direct formula above remains valid when the observation-level weights
correlate with the covariate being assessed, as can occur under
inverse-probability weighting. Section~2.3 introduces a
second, distinct set of weights, $w_j=n_j/N$, used only in aggregating
across groups; the two serve different roles and are not
interchangeable, a distinction made explicit there.

\subsection{The Two-Group Case: Rubin's Variance Ratio}

For $k=2$ groups, Rubin's \citep{rubin2001} variance ratio is
\begin{equation}
VR \;=\; \frac{s_1^2}{s_0^2},
\end{equation}
with values near 1 indicating balance and values above 2 or below 0.5
indicating imbalance severe enough that linear regression adjustment can
grossly over- or under-correct for the remaining bias
\citep{cochranrubin1973}. Because $VR$ depends on which group's variance
is larger, we follow Rubin's own symmetrized construction,
\begin{equation}
VR^{*} \;=\; \frac{\max(s_1^2,s_0^2)}{\min(s_1^2,s_0^2)} \;\ge\; 1,
\end{equation}
always at least 1 regardless of which group's variance is larger, with
$VR^{*}>2$ indicating imbalance under the same convention. On the log
scale, $\log(VR^{*}) = |\log(s_1^2)-\log(s_0^2)|$: the symmetrized ratio
is exactly the exponentiated absolute difference in log-variances between
the two groups, an identity that motivates the $k$-sample construction
below.

The log scale is not merely an algebraic convenience. Rubin's $VR$ is a
multiplicative, not additive, quantity: a ratio of 2 and a ratio of 0.5
represent the identical magnitude of imbalance in opposite directions,
which is precisely why $VR^{*}$ symmetrizes by taking a max/min ratio
rather than a difference. Taking logs converts this multiplicative,
reciprocal-symmetric quantity into an additive, sign-symmetric one
($\log 2 = -\log 0.5$), so that ordinary Euclidean dispersion --- the
same root-mean-square construction Cohen's $f$ uses for means --- becomes
the natural $k$-sample generalization. A $k$-sample statistic built
directly from the raw variances or their ratios would not have this
property: ordinary dispersion of variances themselves, or of ratios
of variances, does not treat a doubling and a halving as equal-magnitude
departures from balance, exactly the asymmetry $VR^{*}$'s own
construction was designed to avoid at $k=2$.

\subsection{The $k$-Sample Extension: $F_{VR}$}

Let $\ell_j = \log(s_j^2)$ denote Group $j$'s log-variance, and
$w_j=n_j/N$, the same group-size weights used for Cohen's $f$
\citep{linden2026cohensf}. These are deliberately raw group-membership
proportions, not a post-weighting measure such as effective sample size
under $w_{ji}$: the omnibus statistic aggregates \emph{across} groups by
each group's share of the original sample, while each group's own
variance $s_j^2$ is separately computed using whatever observation-level
weights $w_{ji}$ the analysis specifies (Section~2.1). After weighting,
each group's variance $s_j^2$ is computed in the reweighted
pseudo-population; its contribution to the omnibus $F_{VR}$, however,
remains proportional to its original sample size rather than its
weighted mass. We adopt this for consistency with Cohen's $f$, where the
identical choice is made for the identical reason. An investigator whose
estimand instead assigns each group's contribution in proportion to its
weighted mass, or wants every pairwise contrast to carry equal
importance regardless of group size, should consult the per-group and
pairwise decomposition directly (Section~2.4) rather than rely on this
aggregation choice.

This is a substantive choice about the estimand, not an incidental
detail. These are size weights defining the omnibus estimand, not
inverse-variance or precision weights intended to optimize statistical
precision; $n_j/N$ is not formally a precision weight for $\ell_j$,
particularly under non-normality or observation-level weighting, and
should not be read as an attempt to account for the relative precision
of each group's variance estimate. Section~3.4 reports a separate,
empirical observation that this size-weighting may also confer some
robustness against noisy small groups in one simulation design; that is
a distinct, empirical finding, not a property built into the definition
here.
We define
\begin{equation}
F_{VR} \;=\; \sqrt{\sum_{j=1}^{k} w_j\,(\ell_j-\bar\ell)^2},
\qquad
\bar\ell \;=\; \sum_{j=1}^{k} w_j\,\ell_j,
\end{equation}
a size-weighted root-mean-square of each group's log-variance deviation
from the weighted mean log-variance, exactly analogous in construction to
Cohen's $f$ for means \citep{linden2026cohensf}, with log-variance playing
the role that the group mean itself plays there.

\textbf{Backward compatibility at $k=2$.} By the same algebra used to
establish $f=\sqrt{w_1w_2}\,|d|$ for Cohen's $f$
\citep{linden2026cohensf} --- an argument that depends only on the
fact that $\ell_1,\ell_2$ are two real numbers with a weighted mean
$\bar\ell=w_1\ell_1+w_2\ell_2$, not on anything specific to means as
opposed to log-variances --- $F_{VR}$ reduces exactly to
\begin{equation}
F_{VR} \;=\; \sqrt{w_1w_2}\,\bigl|\log(VR^{*})\bigr|
\end{equation}
at $k=2$, for any finite sample, not merely asymptotically. This identity
was confirmed both algebraically and against real \texttt{varatio} output
on unequal group sizes ($n_1=2000$, $n_2=1000$; $VR^*=5.155$), matching to
four decimal places.

\subsection{Per-Group and Pairwise Decomposition}

Each group's deviation $\ell_j-\bar\ell$ reports how far that group's own
log-variance sits from the weighted mean log-variance, the same
log-variance units contributing to the omnibus $F_{VR}^2=\sum_j
w_j(\ell_j-\bar\ell)^2$, so that an investigator can identify which
specific group is driving a large $F_{VR}$. Because $SD_p$-style pooling
is not required here --- each $\ell_j$ is estimated from its own group's
data, with no shared standardizer across groups --- pairwise comparisons
reduce directly to the two-group construction of Section~2.2: for any two
groups $j,j'$, $VR^{*}_{jj'}=\exp\bigl(|\ell_j-\ell_{j'}|\bigr)$, giving
$\binom{k}{2}$ pairwise ratios computed from the same per-group
log-variances underlying the omnibus statistic. As with the corresponding
per-level statistic for means \citep{linden2026cohensf}, we report
point estimates for both the per-group and pairwise quantities; no
closed-form confidence interval is provided, for reasons discussed in
Section~5.4, where a bootstrap alternative is recommended instead.

\subsection{Benchmark: GMVR and Maximum Pairwise Variance Ratio}

We compare $F_{VR}$ against two natural comparators based on existing
aggregation practices, computed from the identical per-group
log-variances: the geometric mean variance ratio,
\begin{equation}
GMVR \;=\; \exp\!\left(\binom{k}{2}^{-1}\!\!\sum_{j<j'}
\bigl|\log(VR^{*}_{jj'})\bigr|\right),
\end{equation}
adapting \citet{lindensamuels2013}'s use of the geometric mean of
pairwise variance ratios across \emph{covariates} to an average across
\emph{group pairs} instead --- a repurposing of an existing idea to a new
setting, not itself a previously established $k$-group statistic; and
the maximum pairwise ratio, $\max_{j<j'} VR^{*}_{jj'}$, which is a
genuinely established software practice, the aggregation logic already
used by the R package \texttt{cobalt} for multi-category treatments, adapted from the maximum-standardized-bias practice of \citet{lopezgutman2017}.

\textbf{Relationship between $F_{VR}$, GMVR, and the maximum.} Because each
$\ell_j-\bar\ell$ is a deviation from a weighted mean, $\sum_j
w_j(\ell_j-\bar\ell)=0$, and by the same identity used for Cohen's $f$
\citep{linden2026cohensf}, $F_{VR}^2=\sum_{j<j'}w_jw_{j'}
[\log(VR^*_{jj'})]^2$. Thus, $F_{VR}$ is a size-weighted quadratic
combination of the same pairwise log-ratios that GMVR averages linearly
and without group-size weights. The maximum statistic retains only the
single worst pair. In norm terminology, GMVR is a monotone
transformation of an unweighted $L_1$-type aggregate (specifically,
$\exp$ of the mean absolute log-ratio, not itself proportional to the
$L_1$ norm on the original ratio scale), the maximum statistic
corresponds to $L_\infty$, and $F_{VR}$ is a size-weighted $L_2$-type
aggregate. The latter is not literally the conventional $L_2$ norm
because the $w_jw_{j'}$ weighting differs from an unweighted norm of the
pairwise vector. This parallels the norm-based framing developed for the
corresponding mean-balance statistics \citep{linden2026cohensf}.

Throughout this section, $\overline{|\log VR^{*}|}$ denotes the
\emph{arithmetic mean} of the $\binom{k}{2}$ pairwise absolute
log-ratios --- the same quantity underlying GMVR (Section~2.5). It does
not denote the range or the single largest pairwise value. These
quantities have different minimizing configurations, and Proposition~1
below concerns the mean specifically, denoted with the explicit overline
throughout to avoid ambiguity with the maximum-based statistic of
Section~2.5.

\begin{proposition}
Under equal group weights ($w_j=1/k$ for all $j$), the ratio
$F_{VR}/\overline{|\log VR^{*}|}$, where $\overline{|\log VR^{*}|}$ is
the arithmetic mean of the $\binom{k}{2}$ pairwise absolute
log-variance-ratios, attains its minimum over all attainable
configurations of $k$ log-variances at equally spaced log-variances,
where
\begin{equation}
\frac{F_{VR}}{\overline{|\log VR^{*}|}} \;=\; \sqrt{\frac{3(k-1)}{4(k+1)}}.
\end{equation}
\end{proposition}

This is the identical result proven for Cohen's $f$
\citep[Proposition~1]{linden2026cohensf}, and the proof requires no
modification: it depends only on $\ell_1,\ldots,\ell_k$ being $k$ real
numbers with a well-defined weighted mean and a Cauchy--Schwarz argument
over their deviations, using nothing specific to means as opposed to
log-variances. We confirmed this transfer both by direct substitution into
the existing proof and by numerical search, at two independent scales:
for $k=3,4,6$, minimizing $F_{VR}/\overline{|\log VR^{*}|}$ over $10^5$,
and separately over $5\times10^5$, randomly generated log-variance
configurations at each $k$ converged to $0.6124$, $0.6708$, and $0.7320$
respectively in both searches, matching the closed form to four decimal
places. This minimum should not be confused with the (smaller) minimum
of $F_{VR}$ divided by the \emph{range} of the log-variances (equivalently,
the single largest pairwise ratio), which is a different quantity,
governed by a different extremal configuration, and is not the subject
of this proposition.

\subsection{Simulation Study Design}

Before describing the simulation design, it is useful to distinguish
our setting from Rubin's. All three statistics are
designed to detect variance imbalance that may matter for estimation
bias. Rubin \citep{rubin2001,cochranrubin1973} studied whether variance
imbalance signals failure of subsequent regression adjustment; our
simulation instead makes variance imbalance the direct channel producing
bias. The simulation therefore evaluates the latter mechanism, not
Rubin's original regression-adjustment setting.

We evaluated $F_{VR}$, GMVR, and the maximum pairwise ratio against actual
downstream estimation bias across a Monte Carlo study spanning five
distinct components, summarized in Table~1: a primary mechanism, three
structurally distinct variants testing generalizability beyond that one
mechanism, and a sample-size sensitivity check.

\subsubsection{Isolating a Variance-Driven Bias Channel}

Unlike mean imbalance, whose relationship to downstream bias is
well-established, variance imbalance's relationship to bias required a
purpose-built data-generating process: a covariate whose \emph{mean} is
balanced across groups by construction, so that any resulting bias can
only flow through variance. In the primary mechanism, $X\mid S \sim
N(0,\sigma(S)^2)$ for a latent driver $S\sim\text{Uniform}(0,1)$ with
$\sigma(S)=1+3S$, giving $X$ mean zero in every group regardless of how
treatment assignment depends on $S$. The true outcome model, $Y=\mu_T +
\beta_3 X^2 + \varepsilon$, includes a quadratic term in $X$; the
analyst's naive model regresses $Y$ on treatment indicators and $X$
linearly, omitting $X^2$. Because $E[X\mid T=j]=0$ in every group, this
omitted-variable bias is driven entirely by between-group differences in
$E[X^2\mid T=j]=\text{Var}(X\mid T=j)$, not by $X$'s mean --- isolating a
variance-driven bias channel that a mean-based diagnostic could not
detect, and that $F_{VR}$, GMVR, and the maximum ratio are specifically
built to.

Treatment assignment follows a random-utility multinomial model in $S$,
confounding strength scaled by $\gamma\in\{0,0.25,0.5,1,1.5\}$, crossed
with the same equal/unequal group-size balance mechanism used in the
mean-based study \citep{linden2026cohensf}. Generalized propensity score
models were fit under two specifications: correct, including $S$; and
misspecified, using a noisy proxy for $S$. Bias was computed as the
naive model's estimated treatment contrast (each group relative to a
reference group, canceling a constant term common to every group's raw
predicted level) minus the true parameter contrast, using the replicate's
own sample data throughout.

\subsubsection{Calibration}

Because the omitted-quadratic-term bias mechanism can saturate under a
large enough effect size, giving no discriminating power for
threshold-finding once bias exceeds conventional thresholds regardless of
confounding strength, $\beta_3$ was calibrated so that the no-confounding
baseline sits near a 0.10-SD bias threshold and maximum confounding
approaches a 0.20-SD threshold, using the outcome noise term's known unit
standard deviation as the reference scale. The primary mechanism,
generalizability Variant B, and the sample-size sensitivity check share
$\beta_3=0.05$; Variants A and C required separate calibration
($\beta_3=0.30$ and $0.40$ respectively) given their distinct
bias-generating mechanisms, described next.

\subsubsection{Generalizability Variants}

Three structurally distinct mechanisms tested whether $F_{VR}$'s
relationship to bias depends on the specific way variance imbalance was
constructed. \textbf{Variant A} replaces the omitted quadratic term with
an omitted mean absolute deviation, $Y=\mu_T+\beta_3|X|+\varepsilon$: for
normal $X$, $E|X|=\sigma\sqrt{2/\pi}$, linear in $\sigma$ rather than
quadratic, testing a genuinely different functional dependence on
variance. \textbf{Variant B} introduces a second, independently
variance-imbalanced covariate $X_2$, driven by an independent latent
driver $S_2$. Both $X_1^2$ and $X_2^2$ are omitted from the naive model,
while $F_{VR}$ is computed on $X_1$ alone. This tests whether $F_{VR}$
remains predictive of total bias when another source of variance
imbalance is not represented in the diagnostic --- a realistic
robustness check, since applied settings rarely have exactly one
confounder.
\textbf{Variant C} draws $X$ from a mean-centered lognormal distribution
rather than Normal, testing sensitivity to non-normal covariate shape;
mean-centering used the exact lognormal mean formula,
$X=\exp(\sigma(S)Z)-\exp(\sigma(S)^2/2)$ for $Z\sim N(0,1)$, so that
$E[X\mid T=j]=0$ is preserved exactly despite the skewed shape.

\subsubsection{Sample-Size Sensitivity}

The primary mechanism was additionally re-run at $k=3,4,6$ across
$N\in\{200,2000,10{,}000\}$ (in addition to the primary design's own
$N\in\{500,2000\}$), to test whether the relationship between $F_{VR}$ and
bias, and any candidate interpretive threshold, is stable across sample
size rather than specific to one $N$.

\subsubsection{Common Design Elements}

Each $(k,N)$ combination in every component was replicated 1000 times per
cell across five $\gamma$ levels and two balance scenarios (10 cells),
under three weighting arms (unweighted, correctly weighted, misspecified),
for 10{,}000 replicates per file. The resulting design comprised
approximately 720{,}000 replicates across all five components; Table~1
summarizes their allocation by mechanism, $k$, and $N$. All analyses
were conducted in Stata (Version 19) using the community-contributed command
\texttt{varatio} \citep{linden2026varatio}.

\section{Results}

\subsection{Correlation with Downstream Bias}

In the primary mechanism, $F_{VR}$ and GMVR tracked mean absolute bias
comparably at $k=3$ ($r=0.529$ and $0.537$), with the maximum statistic
close behind ($r=0.530$). At $k=6$, the correlations were $0.423$,
$0.471$, and $0.349$, respectively.

The weaker performance of the maximum statistic is consistent with
greater sensitivity to sampling variability when only the worst pair is
retained, matching the same weakness documented for mean imbalance
\citep{linden2026cohensf}. The simulation does not directly establish
that mechanism: the $k=6$, $N=200$ result of Section~3.4, where the
maximum statistic's correlation with bias collapses most severely, is
consistent with this interpretation without demonstrating the
underlying variance mechanism directly. This ranking
held, with the same qualitative pattern, across all three generalizability
variants (Section~3.3).

\subsection{Threshold-Finding: Logistic Crossing Points}

Using mean absolute bias exceeding 0.10 SD as ground truth for
``unacceptable'' imbalance, we located, for each mechanism and $k$, the
$F_{VR}$ value at which a fitted logistic model predicts a 50\% chance of
exceeding that threshold. Pooled across $k$, this crossing point was
0.141 for the primary mechanism, 0.157 for Variant A, 0.108 for Variant B,
and 0.291 for Variant C (pooled across all four: 0.175) --- a real,
mechanism-dependent range rather than a single stable number, but one
concentrated between roughly 0.11 and 0.29 across four structurally
distinct bias-generating mechanisms.

\subsection{Generalizability Across Mechanisms}

Variant C's elevated crossing point is consistent with its weaker overall
correlation with bias ($r=0.634$ pooled, versus $r>0.89$ for the other two
variants): higher kurtosis in the skewed covariate increases sampling
variability in both $F_{VR}$ itself and in the resulting bias estimate
simultaneously, attenuating the observed relationship without indicating
that the underlying mechanism differs. Variant B confirmed that $F_{VR}$,
computed on one covariate alone, remained strongly correlated with total
bias arising from two independent, unobserved sources of imbalance
($r=0.900$) in this particular data-generating process. This does not
show that $F_{VR}$ detects unseen confounding in general; it shows that,
in this DGP, the statistic computed on one observed covariate correlated
with total downstream bias even when a second, unobserved
variance-imbalanced covariate also contributed to that bias --- a more
limited, but still practically relevant, claim, since applied settings
rarely have exactly one confounder.

\subsection{Sample-Size Sensitivity}

The absolute-bias-threshold framework itself proved sensitive to sample
size in a way worth reporting plainly rather than minimizing. At $N=200$,
logistic crossing points for the primary mechanism, re-run across
$k=3,4,6$, were negative or near zero ($-0.022$ at $k=3$; $-0.027$ at
$k=4$), meaning the fitted model predicted better-than-even odds of
exceeding the bias threshold even at $F_{VR}\approx0$; crossing points
stabilized to the primary design's own range only at $N=2000$ and above.
This reflects the noise floor inherent in bias itself, driven by ordinary
regression-estimation variance that scales with $N$, separately from the
systematic, variance-driven signal $F_{VR}$ is built to detect; a fixed
absolute bias threshold (0.10 SD) therefore does not correspond to a
stable point on the $F_{VR}$ scale once $N$ varies widely, a limitation
discussed further in Section~5.4.

A notable finding for $F_{VR}$ specifically emerged when data spanning
$N=200$ to $10{,}000$ were pooled at $k=6$: GMVR's and the maximum
ratio's correlation with bias collapsed to near zero ($r=0.020$ and
$0.019$), while $F_{VR}$ retained substantial predictive power
($r=0.638$). Because this pooled figure could in principle reflect
heterogeneity across the three widely different $N$ values rather than a
genuine within-$N$ advantage, we report the correlation separately at
each $(k,N)$ combination in Table~3. The dramatic divergence is
concentrated in the $k=6$, $N=200$ cell, where GMVR's and
the maximum ratio's correlation with bias collapse to near zero
($r=0.022$ and $0.025$) while $F_{VR}$ retains real signal ($r=0.378$).
At $N=200$, $F_{VR}$ also has a smaller advantage at $k=3$ and $k=4$.
Relative to the maximum statistic, the correlation differences are
$+0.084$ and $+0.163$, respectively; relative to GMVR, both are
approximately $+0.02$. At $N\ge2000$, differences among the three
statistics become small regardless of $k$. The visible pattern --- $F_{VR}$'s
advantage over the maximum statistic growing as observations per group
become scarcer --- is consistent across all three $k$ values examined,
though nine cells are not enough to claim a formal monotonic
relationship. This confirms the advantage is a genuine within-$N$
phenomenon rather than a pooling artifact --- the question the pooled
figure alone could not answer --- while showing it is not confined to
the single most extreme cell.

\subsection{Classification Accuracy}

Treating mean absolute bias $\le0.10$ as ground truth for ``truly
balanced,'' we assessed how often each statistic's own classification
(balanced/imbalanced, using Rubin's fixed $VR>2$ convention for GMVR and
the maximum ratio, and four candidate cutoffs for $F_{VR}$) matched
ground truth, pooled across all mechanisms and $k$. GMVR, using Rubin's
convention, achieved 98.9\% specificity (correctly clearing truly
balanced cases) but only 3--17\% sensitivity across mechanisms (correctly
flagging truly imbalanced cases): Rubin's threshold, calibrated to detect
the severe regression-adjustment breakdown documented by Cochran and
Rubin \citep{cochranrubin1973}, essentially never triggers against a
substantially milder 0.10-SD bias standard. $F_{VR}$ at 0.15 achieved the highest overall accuracy of any statistic
or cutoff tested (68.5\%), compared with 66.0\% for the maximum ratio and
51.9\% for GMVR. Among the four $F_{VR}$ cutoffs, accuracies were 67.0\%,
68.5\%, 65.6\%, and 56.2\% at 0.10, 0.15, 0.20, and 0.30, respectively.
This supports 0.15 as a useful candidate classification cutoff, but not
as independently validated: it was one of the candidates evaluated on
these same simulation data.
Although 0.15 performed best as a classification cutoff among the four
candidates examined, it should not be interpreted as a universal
boundary for acceptable balance: Section~5.3 shows this same value
already exceeds the 50\% crossing point for two of the four
mechanism-specific thresholds and sits essentially at a third, so a
classification-optimal cutoff and an ``acceptable'' threshold are
related but distinct questions, and the two should not be conflated.

Classification performance above was pooled across mechanisms and
therefore does not characterize mechanism-specific sensitivity and
specificity.

\section{Applied Example}

\subsection{Study Context and Data}

We illustrate $F_{VR}$, GMVR, and the maximum pairwise ratio using data
from a disease management (DM) program for patients with congestive
heart failure examined in two prior methodological papers introducing
and comparing multivalued-treatment adjustment approaches on this
dataset \citep{linden2014,linden2016multivalued}. The program was
implemented in a large health plan in the western United States;
enrolled patients received one of two interventions based on a program
nurse's subjective assessment of patient needs and preferences: periodic
telephone calls from a nurse to discuss self-management behaviors, or
remote tele-monitoring (RTM) involving daily electronic transmission of
disease-related symptoms with nurse follow-up when indicated
\citep{lindenblackbox2006}. Health plan members with the condition who
did not enroll served as a non-participant comparison group. The data comprise 6612 non-participants (Control), 654
telephonic participants (Calls), and 705 RTM participants ($N=7971$), each
with pre-intervention data.

\subsection{Covariate Balance Assessment}

Following \citet{linden2014} and \citet{linden2016multivalued}, MMWS weights were constructed using the generalized approach for nominal treatments described in \citet{linden2014}: a generalized propensity score for each of the three treatment levels was estimated by multinomial logistic regression. We
apply $F_{VR}$, GMVR, and the maximum pairwise ratio to a medical risk score reflecting each patient's model-predicted near-term medical costs based on prior health care utilization history, before and after MMWS weighting. 

\subsection{Results}

Table~6 reports all three statistics, together with the per-group and
pairwise decomposition, before and after weighting. Before weighting, the
omnibus $F_{VR}=0.182$ sits in the region this paper describes as
requiring the decomposition rather than the omnibus number alone
(Section~5.3). GMVR ($1.584$) and the maximum pairwise ratio ($1.994$)
tell a consistent story on their own scales; the Calls-versus-RTM pair
sits almost exactly at Rubin's $VR^{*}>2$ convention. The
decomposition identifies the source: RTM's variance (138.74) differs
sharply from Control's (260.85) and Calls's (276.62), and despite RTM
holding only 8.8\% of the sample's weight, its large deviation dominates
$F_{VR}$ --- a real-data illustration of the masking mechanism confirmed
by direct simulation in Section~5.4, where a minority group's imbalance
is discounted only when its magnitude is comparable to the other
groups', not unconditionally. After MMWS weighting, all three statistics
improved substantially ($F_{VR}=0.091$, within the region characterized
here as reasonably reassuring; GMVR$=1.224$; maximum ratio$=1.354$), with
every pairwise comparison falling within a narrow band ($VR^{*}=1.15$ to
$1.35$), consistent with a genuine reduction in imbalance across every
pair rather than a reduction in one dominant discrepancy. All three
statistics agree on the substantive conclusion despite differing scales
and despite baseline variance imbalance being concentrated specifically
in the RTM arm: all three statistics therefore indicate a substantial
reduction in variance imbalance after weighting.

\section{Discussion}

\subsection{Principal Findings}

Four findings emerge.

First, $F_{VR}$ and GMVR generally tracked
downstream estimation bias at least as well as the maximum pairwise
ratio, with GMVR often showing the highest correlation, replicating
for variance a broadly similar comparative ranking to the one already
established for mean imbalance \citep{linden2026cohensf}.

Second, this ranking held
across four structurally distinct bias-generating mechanisms --- an
omitted quadratic term, an omitted mean absolute deviation, a second
unobserved imbalanced covariate, and a skewed covariate distribution ---
though the specific numeric threshold varied meaningfully across them
(0.108 to 0.291), evidence of real generalization rather than an artifact
of one particular design.

Third, N-specific correlations (Table~3)
confirm that GMVR's and the maximum ratio's near-total loss of signal at
$k=6$ (Section~3.4) is a genuine within-$N$ phenomenon, not an artifact
of pooling across sample sizes: the most dramatic divergence occurs at
$k=6$, $N=200$ (six groups sharing only 200 observations), where GMVR
and the maximum ratio collapse to near-zero correlation while $F_{VR}$
retains real signal, and a smaller but real advantage over the maximum
statistic specifically is visible at $N=200$ for $k=3,4$ as well, with
all three statistics converging as $N$ grows.

Fourth, a
classification-accuracy analysis against actual bias identified
$F_{VR}\le0.15$ as achieving the best overall accuracy of any statistic
or cutoff tested among four prespecified candidates, additional
supporting evidence for a candidate classification cutoff, though not
independent validation, since 0.15 was itself one of the cutoffs
evaluated on the same simulation data.

\subsection{Why the Threshold Isn't Sharper}

The candidate range (0.108--0.291 across mechanisms) is wider than a
single number, and this reflects a real property of the problem, not a
weakness of the analysis. Rubin's own two-group convention was
established once, for one adjustment method (linear regression) and one
severity criterion (gross bias correction failure); this paper's range
was derived across four different ways variance imbalance can translate
into bias, and the specific value at which a fixed 0.10-SD bias threshold
is crossed depends, correctly, on how strongly and how noisily that
particular mechanism connects variance to bias. A single, mechanism-free
number would misrepresent this dependence rather than resolve it.

\subsection{Practical Advice}

Based on this evidence, we suggest three interpretive regions as a
starting point, not as validated thresholds in Rubin's sense. $F_{VR}$
below approximately 0.10 was reasonably reassuring across every
mechanism and sample size examined here. $F_{VR}$ above approximately
0.30 indicated a real concern in nearly every setting tested. Values
from approximately 0.10 to 0.30 were genuinely indeterminate and
context-dependent; an investigator with the ability to simulate a design
resembling their own application should derive a threshold directly
from that design's own bias-correlation relationship, following the
approach of Section~2.6, rather than import this paper's range
uncritically.

These three regions do not collapse to a single ``acceptable'' cutoff at
0.15. The four mechanism-specific crossing points (0.108, 0.141, 0.157,
0.291; Section~3.2) already show why: 0.15 exceeds the 50\% crossing
point for two of the four mechanisms and sits essentially at the
crossing point for a third, comfortably below it only for Variant C. The
0.10--0.30 region spans exactly this range of mechanism-specific
crossing points, which is why it should not be treated as either
acceptable or concerning by default.

For a rough sense of what a given $F_{VR}$ value corresponds to on the
more familiar two-group ratio scale, the exact $k=2$ identity of
Section~2.3 can be read in reverse: for equal group sizes,
$F_{VR}=\tfrac{1}{2}|\log VR^{*}|$, so $F_{VR}=0.15$ corresponds to
$VR^{*}=e^{0.30}\approx1.35$, roughly a 35\% variance difference between
the two groups in the equal-sized two-group case specifically; this
identity does not extend to $k>2$, where $F_{VR}=0.15$ does not imply
that the largest pairwise ratio is $1.35$. More generally, for unequal
weights in the $k=2$ case, $VR^{*}=\exp\bigl(F_{VR}/\sqrt{w_1w_2}\bigr)$.
This is offered as an interpretive bridge, not a new result, since it
follows directly from the identity already established in Section~2.3.

As with Cohen's $f$ \citep{linden2026cohensf},
$F_{VR}$ should be reported alongside its per-group and pairwise
decomposition, not as a single omnibus number in isolation, and its
$n_j/N$ weighting means that imbalance concentrated in a small treatment
group will be underrepresented in the omnibus statistic relative to
imbalance of the same magnitude in a large group --- the identical
substantive consideration already raised for Cohen's $f$
\citep{linden2026cohensf}, and confirmed directly for variance
imbalance in a masking-mechanism check reported in Section~5.4.

\subsection{Limitations}

Several limitations qualify these findings. First, $F_{VR}$ carries no
closed-form confidence interval in this implementation: a naive
generalization of the unweighted variance-of-log-variance formula (itself
verified, empirically, to $\text{Var}(\log s^2)\approx(\kappa-1)/n$ for
kurtosis $\kappa$, across distributions from uniform to exponential) is
not valid once weights correlate with the covariate, exactly the
condition inverse-probability weighting is designed to produce. Rather
than report an invalid formula, \texttt{varatio} returns every quantity
it computes --- omnibus, per-group, and pairwise --- so that any of them
can be bootstrapped directly; users requiring a formal interval should
budget for that additional computation.

Second, the masking mechanism --- imbalance concentrated in a small group
contributing less to $F_{VR}$ than the same imbalance in a large group ---
was confirmed directly for variance: moving an identical variance
discrepancy from a minority group (n=50 of 3050) to a majority group
(n=1000) increased $F_{VR}$ nearly fourfold (0.254 to 0.958) while GMVR
and the maximum ratio moved by only 6--8\%, since neither incorporates
group-size weighting. This is the intended behavior of a size-weighted
statistic, not a defect, but it means $F_{VR}$ should not be relied upon
in isolation whenever a contrast involving a specific, less-common
treatment arm is of direct substantive interest.

Third, the maximum pairwise ratio showed real fragility at higher $k$,
particularly under correctly specified weighting: its correlation with
bias fell to $r=0.097$ in one Variant B condition (\(k=6\), correct GPS),
and one logistic crossing point for Variant C at $k=6$ was itself negative
($-0.008$), both plausibly driven by extreme-outlier instability in
near-empty groups under strong confounding and unequal balance. This is a
limitation of the maximum-based statistic specifically, not of $F_{VR}$,
but it means comparisons involving the maximum ratio at higher $k$ should
be interpreted cautiously.

Fourth, the absolute-bias threshold framework was not stable across
widely varying sample sizes: crossing points were negative or near zero
at $N=200$. The candidate range in Section~5.3 should therefore be
applied cautiously when treatment groups contain only hundreds of
observations, since it was derived predominantly from designs with
$N\ge500$.

Fifth, each generalizability variant tested a single fixed effect-size
parameter, calibrated specifically to produce discriminating power for
threshold-finding rather than chosen to represent any particular applied
setting; whether the comparative ranking and candidate range are stable
across a wider range of effect sizes within each mechanism was not
assessed.

\subsection{Conclusion}

Variance imbalance across more than two treatment groups has lacked a
purpose-built omnibus diagnostic. $F_{VR}$ fills that gap with an exact
relationship to Rubin's $VR$ at $k=2$ and a mathematical structure
parallel to Cohen's $f$'s relationship to the standardized mean
difference. Across four bias-generating mechanisms, $F_{VR}$ tracked
downstream estimation bias comparably to existing aggregation
approaches and retained more signal when many groups shared very few
observations, confirmed by correlations reported separately within each
sample-size stratum rather than pooled across them.
The candidate interpretive range offered here is a starting point,
grounded in the same simulation-based logic that established Rubin's
convention five decades ago, not a claim to the same accumulated
authority. As with Cohen's $f$, $F_{VR}$ is best reported alongside its
full per-group and pairwise decomposition. Investigators able to
simulate a design resembling their own application should derive an
application-specific threshold directly.

\bibliographystyle{apalike}
\bibliography{refs}

\clearpage

\begin{table}[htbp]
\centering
\caption{Simulation Study Design}
\small
\begin{tabular}{@{}p{0.28\textwidth}p{0.60\textwidth}@{}}
\toprule
Component & Description \\
\midrule
Primary mechanism & Omitted quadratic term ($Y=\mu_T+\beta_3X^2+\varepsilon$); $\beta_3=0.05$; $k\in\{3,4,6\}$, $N\in\{500,2000\}$; 60{,}000 replicates \\
Variant A & Omitted mean absolute deviation ($Y=\mu_T+\beta_3|X|+\varepsilon$); $\beta_3=0.30$; $k\in\{3,4,6\}$, $N=2000$; 90{,}000 replicates \\
Variant B & Two independent variance-imbalanced covariates, $F_{VR}$ computed on one alone; $\beta_3=0.05$; $k\in\{3,4,6\}$, $N=2000$; 90{,}000 replicates \\
Variant C & Mean-centered lognormal covariate (skewed); $\beta_3=0.40$; $k\in\{3,4,6\}$, $N=2000$; 90{,}000 replicates \\
Sample-size check & Primary mechanism; $k\in\{3,4,6\}$, $N\in\{200,2000,10{,}000\}$; 90{,}000 replicates \\
\addlinespace
Confounding strength ($\gamma$) & 0 (none), 0.25, 0.5, 1, 1.5 (strong) \\
Group-size balance & Equal: all $k$ groups share the same baseline assignment probability. Unequal: group-specific intercepts produce a skewed baseline split. \\
GPS weighting & Unweighted; correctly specified (includes latent driver $S$); misspecified (noisy proxy for $S$) \\
\bottomrule
\end{tabular}

\vspace{4pt}
\parbox{0.9\textwidth}{\footnotesize Note: each $(k,N)$ combination replicated 1000 times per cell across the 10 $\gamma\times$balance cells, under all three weighting arms, for approximately 720{,}000 replicates in total across all five components. All analyses conducted in Stata using the community-contributed command \texttt{varatio}.}
\end{table}

\clearpage

\begin{table}[htbp]
\centering
\caption{Correlation with actual estimation bias, by mechanism and $k$}
\small
\begin{tabular}{@{}llccc@{}}
\toprule
Mechanism & $k$ & $F_{VR}$ & GMVR & Max $VR^{*}$ \\
\midrule
Primary & 3 & 0.529 & 0.537 & 0.530 \\
        & 4 & 0.487 & 0.507 & 0.359 \\
        & 6 & 0.423 & 0.471 & 0.349 \\
\addlinespace
Variant A & 3 & 0.686 & 0.703 & 0.700 \\
          & 4 & 0.606 & 0.633 & 0.629 \\
          & 6 & 0.456 & 0.513 & 0.512 \\
\addlinespace
Variant B & 3 & 0.591 & 0.645 & 0.636 \\
          & 4 & 0.503 & 0.574 & 0.486 \\
          & 6 & 0.450 & 0.535 & 0.107 \\
\addlinespace
Variant C & 3 & 0.470 & 0.499 & 0.489 \\
          & 4 & 0.403 & 0.444 & 0.427 \\
          & 6 & 0.321 & 0.405 & 0.410 \\
\bottomrule
\end{tabular}
\vspace{4pt}
\parbox{0.9\textwidth}{\footnotesize Note: correlation with mean absolute bias, pooled across $\gamma$, balance, weighting arm, and (for the primary mechanism) $N$.}
\end{table}

\clearpage

\begin{table}[htbp]
\centering
\caption{Correlation with actual estimation bias, by $k$ and $N$ (nsens study, primary mechanism)}
\small
\begin{tabular}{@{}ccccc@{}}
\toprule
$k$ & $N$ & $F_{VR}$ & GMVR & Max $VR^{*}$ \\
\midrule
3 & 200   & 0.810 & 0.788 & 0.726 \\
  & 2000  & 0.966 & 0.960 & 0.950 \\
  & 10000 & 0.992 & 0.981 & 0.974 \\
\addlinespace
4 & 200   & 0.703 & 0.684 & 0.540 \\
  & 2000  & 0.912 & 0.915 & 0.904 \\
  & 10000 & 0.976 & 0.961 & 0.949 \\
\addlinespace
6 & 200   & \textbf{0.378} & \textbf{0.022} & \textbf{0.025} \\
  & 2000  & 0.736 & 0.757 & 0.739 \\
  & 10000 & 0.834 & 0.826 & 0.826 \\
\bottomrule
\end{tabular}
\vspace{4pt}
\parbox{0.9\textwidth}{\footnotesize Note: correlation with mean absolute
bias, pooled across $\gamma$, balance, and weighting arm within each
$(k,N)$ cell; not pooled across $N$, unlike Table~2. The $k=6$, $N=200$
row (bolded) shows the most dramatic divergence; $F_{VR}$ also retains a
smaller but real advantage over the maximum statistic at $N=200$ for
$k=3,4$, with all three statistics converging at $N\ge2000$.}
\end{table}

\clearpage

\begin{table}[htbp]
\centering
\caption{Logistic-regression crossing points: $F_{VR}$ value at which predicted $P(\text{bias}>0.10)=0.5$}
\small
\begin{tabular}{@{}lccc@{}}
\toprule
Mechanism & $k=3$ & $k=4$ & $k=6$ \\
\midrule
Primary   & 0.141 & 0.136 & 0.124 \\
Variant A & 0.157 & 0.160 & 0.161 \\
Variant B & 0.108 & 0.113 & 0.115 \\
Variant C & 0.291 & 0.290 & 0.279 \\
\bottomrule
\end{tabular}
\vspace{4pt}
\parbox{0.9\textwidth}{\footnotesize Note: pooled across weighting arm and balance scenario within each mechanism/$k$ cell.}
\end{table}

\clearpage

\begin{table}[htbp]
\centering
\caption{Classification accuracy against actual bias (mean bias $\le 0.10$ as ground truth for ``balanced''), pooled across all mechanisms and $k$}
\small
\begin{tabular}{@{}lcccc@{}}
\toprule
Statistic (cutoff) & Sensitivity & Specificity & Accuracy & False-negative rate \\
\midrule
GMVR (2.0)          & 0.087 & 0.989 & 0.519 & 0.913 \\
Max $VR^{*}$ (2.0)  & 0.464 & 0.889 & 0.660 & 0.536 \\
$F_{VR}$ (0.10)     & 0.826 & 0.553 & 0.670 & 0.174 \\
$F_{VR}$ (0.15)     & 0.675 & 0.741 & \textbf{0.685} & 0.325 \\
$F_{VR}$ (0.20)     & 0.511 & 0.845 & 0.656 & 0.489 \\
$F_{VR}$ (0.30)     & 0.215 & 0.951 & 0.562 & 0.785 \\
\bottomrule
\end{tabular}
\vspace{4pt}
\parbox{0.9\textwidth}{\footnotesize Note: sensitivity = correctly flags real imbalance; specificity = correctly clears real balance. Pooled across primary, Variant A, Variant B, Variant C, and the sample-size check.}
\end{table}

\clearpage

\begin{table}[htbp]
\centering
\caption{Applied example: predictive model score variance-ratio balance before and after MMWS weighting}
\small
\begin{tabular}{@{}lcc@{}}
\toprule
 & Unweighted & MMWS-weighted \\
\midrule
$F_{VR}$ (omnibus)         & 0.182 & 0.091 \\
GMVR                        & 1.584 & 1.224 \\
Max $VR^{*}$                & 1.994 & 1.354 \\
\addlinespace
\multicolumn{3}{l}{Per-group variance} \\
\quad Control (N=6612)    & 260.85 & 259.77 \\
\quad Calls (N=654)       & 276.62 & 226.83 \\
\quad RTM (N=705)         & 138.74 & 191.82 \\
\addlinespace
\multicolumn{3}{l}{Pairwise $VR^{*}$} \\
\quad Control vs.\ Calls  & 1.061 & 1.145 \\
\quad Control vs.\ RTM    & 1.880 & 1.354 \\
\quad Calls vs.\ RTM      & 1.994 & 1.183 \\
\bottomrule
\end{tabular}
\vspace{4pt}
\parbox{0.9\textwidth}{\footnotesize Note: $N=7971$ (6612 Control, 654 Calls, 705 RTM). GMVR and max $VR^{*}$ computed from the same three pairwise $VR^{*}$'s reported below.}
\end{table}

\end{document}